\def\uselong{}
\documentclass[conference,10pt,letterpaper]{IEEEtran}

\usepackage{amsmath}
\usepackage{amssymb}
\usepackage{cite}
\usepackage{booktabs}
\usepackage{balance}
\usepackage{graphicx}
\usepackage[pass]{geometry}
\usepackage{placeins}
\usepackage[hidelinks]{hyperref}

\title{Communication-Efficient Model-Distributed Inference for CBRS Radar Detection at the Edge}
\author{\IEEEauthorblockN{Teng Li and Hulya Seferoglu}
\IEEEauthorblockA{Department of Electrical and Computer Engineering, University of Illinois Chicago, Chicago, IL, USA\\
Emails: tli81@uic.edu, hulya@uic.edu}}

\newcommand{\system}{CEDAR-MDI}
\newcommand{\cedaryolo}{CEDAR-YOLO}
\newcommand{\cbrs}{CBRS}
\newcommand{\mdi}{MDI}

\ifdefined\uselong
\newcommand{\appx}[1]{Appendix~\ref{app:#1}}
\else
\newcommand{\appx}[1]{Appendix~#1 of~\cite{cedarmdi2026long}}
\fi

\begin{document}
\bstctlcite{IEEEexample:BSTcontrol}
\maketitle

\begin{abstract}
The Citizens Broadband Radio Service (CBRS) shares the 3550--3700~MHz band between incumbent radar systems and commercial LTE/5G users, so radar activity must be detected accurately and promptly. Spectrogram-based YOLO detectors can localize coexisting radar and cellular signals, but running them on resource-constrained edge devices is challenging. Model-distributed inference (MDI) can increase throughput by partitioning a model across edge devices. Using an end-to-end pipeline from raw I/Q samples to on-device detection, we show that partitioning recent YOLO generations incurs high activation-transfer costs due to their multi-scale structures. To reduce these costs, we propose CEDAR, a communication-efficient, edge-distributed architecture for radar detection. Its RF-aware detector, CEDAR-YOLO, removes YOLO's multi-scale neck and two of its three detection scales, leaving a single-scale path with independently configured frequency and time resolutions, which lowers the communication cost of partitioning. Its compression-aware MDI framework, CEDAR-MDI, integrates lightweight activation compressors and selects, for each compressor, the partition that minimizes the pipeline bottleneck. On an NVIDIA Jetson Orin Nano testbed with real-world over-the-air data, CEDAR-YOLO reduces the parameter count by 35--57\% and model inference time by 16--33\%, and CEDAR-MDI achieves up to \(3.27\times\) the throughput of single-device inference, with detection quality preserved in both cases.
\end{abstract}

\section{Introduction}

The Citizens Broadband Radio Service (\cbrs{}) opens the 3550--3700~MHz band to shared use under three tiers: Incumbent Access, Priority Access, and General Authorized Access. Because incumbent radar systems have the highest priority, commercial transmissions must be suspended or reconfigured whenever radar is detected. Detection must therefore be accurate and timely: missed or delayed detections risk harmful interference to incumbents, whereas false alarms needlessly restrict commercial access.

Radar detection is difficult because radar pulses are short and intermittent, can be weak at the sensor, and may be masked by stronger LTE, 5G, or unauthorized transmissions. Traditional matched filtering and rule-based methods require prior knowledge of the radar waveform and degrade under strong interference and signal overlap~\cite{tassie2023cbrs,soltani2022waldo,reusmuns2024senseoran}.

Object detection on spectrograms offers a promising alternative: it converts I/Q samples into time-frequency images and marks each signal and its occupied region with a class-labeled bounding box. The one-stage You Only Look Once (YOLO) detector balances speed and accuracy and needs no prior knowledge of the radar waveform~\cite{soltani2022waldo}. Existing YOLO-based \cbrs{} work, however, uses early YOLO generations. An over-the-air (OTA) study with YOLOv3 reported a radar recall of only 86.6\%~\cite{tassie2023cbrs}, whereas later generations are both more accurate and faster~\cite{ramos2025yolo}. Recent YOLO-based spectrum-sensing demonstrations~\cite{mathworks2026yoloxcbrs,agarwal2026ettusspectrum} use earlier versions on GPUs rather than on general edge devices without hardware accelerators.

Lightweight YOLO designs, studied extensively for edge devices~\cite{mittal2024lightweight}, reduce the cost of each inference. Model-distributed inference (\mdi{}) instead partitions the model across devices that exchange intermediate activations, distributing storage and computation and pipelining successive inputs to increase throughput~\cite{li2023armdi,li2023msmdi,li2025pamdi,communication_aware_inference_2026}. Data-distributed inference, in which every worker stores the whole model and receives entire inputs, is inefficient at the edge for large inputs~\cite{li2023armdi}. Partitioning YOLO, however, remains sparsely studied and is costly. YOLO's multi-scale feature fusion, skip connections, and multiple detection branches~\cite{ramos2025yolo} mean that a partition boundary can cut through several feature tensors, all of which must be transmitted, which limits the benefit of \mdi{}.

To address these challenges, we present CEDAR (Communication-Efficient, Edge-Distributed Architecture for Radar detection), which reduces the communication cost of \mdi{} at both the model and the system level. Our contributions are as follows:

\begin{itemize}
\item We build an end-to-end pipeline for YOLO-based \cbrs{} detection from raw I/Q samples to on-device deployment, and evaluate recent YOLO generations~\cite{ramos2025yolo,jocher2026yolo26} on real-world OTA data~\cite{tassie2023cbrs}. On an NVIDIA Jetson testbed, we show that their multi-scale structures incur high activation-transfer costs under \mdi{}.
\item We propose \cedaryolo{}, an RF-aware detector designed for \mdi{}. In the spirit of structured pruning, it removes YOLO's multi-scale neck and two of its three detection scales, leaving a single-scale path with independently configured frequency and time resolutions. Relative to the corresponding YOLO models, it reduces the parameter count by 35--57\% and model inference time by 16--33\% while preserving detection quality, and it lowers the communication cost of partitioning.
\item We design \system{}, a compression-aware \mdi{} framework. It integrates lightweight activation compressors, including top-k sparsification, INT8/INT4 quantization, and an LLM.int8()-style hybrid~\cite{dettmers2022llmint8}. For each compressor, it selects offline the partition that minimizes the pipeline bottleneck based on profiled costs. It achieves up to \(3.27\times\) the throughput of single-device inference while preserving detection quality.
\end{itemize}

Our code and trained models are open source~\cite{cedarmdi2026code}, and \ifdefined\uselong the appendix\else the extended version~\cite{cedarmdi2026long}\fi{} provides implementation details and additional results.

\section{YOLO-Based CBRS Detection Pipeline}
\label{sec:yolo_detector_pipeline}

\subsection{Dataset and Metrics}
\label{sec:dataset}

\begin{figure}[!t]
\centering
\includegraphics[width=\columnwidth]{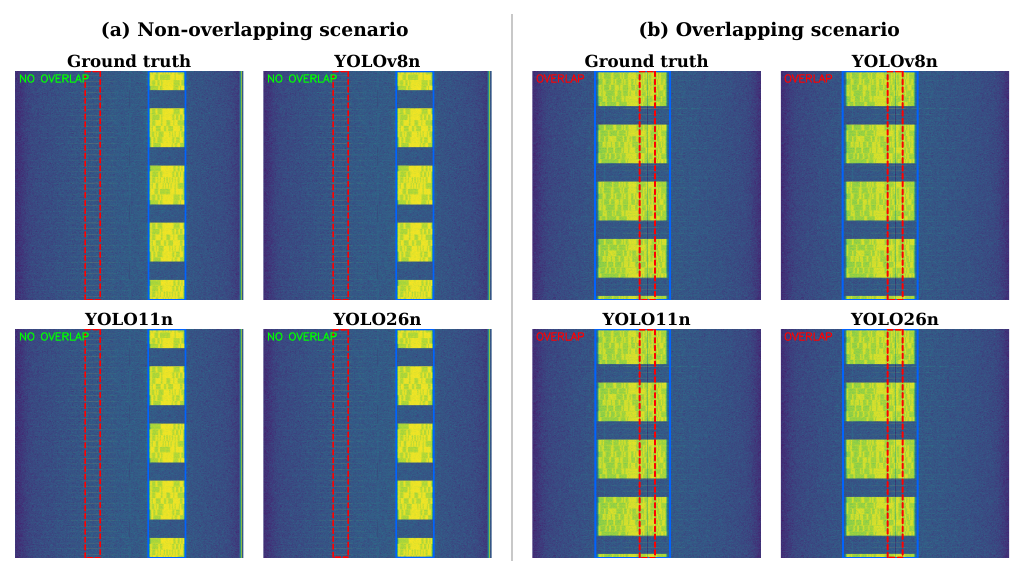}
\caption{Ground truth and predictions of YOLOv8n, YOLO11n, and YOLO26n for non-overlapping and overlapping frames. Blue solid boxes mark LTE, and red dashed boxes mark radar. Overlap is determined by their frequency intervals.}
\label{fig:cbrs_detection_examples}
\end{figure}

We use the real-world OTA dataset of~\cite{tassie2023cbrs}, which contains 5,640 frames. Each frame records 40~ms of I/Q samples with an LTE signal of 5--20~MHz and a 2-MHz radar signal at signal-to-interference-plus-noise ratios (SINRs) from 15 to 35~dB. We split the frames 70/15/15 into training, validation, and test sets within each LTE bandwidth, and the 848 test frames contain 558 overlapping and 290 non-overlapping cases. Each frame is converted into a \(640\times640\) log-magnitude spectrogram, with frequency on the horizontal axis and time on the vertical axis, in which the detector localizes the LTE and radar signals with class-labeled boxes (Fig.~\ref{fig:cbrs_detection_examples}). A frame is overlapping when an LTE box and a radar box intersect along the frequency axis (\appx{A}). We measure box detection by the mean average precision at an intersection-over-union threshold of 0.5 (mAP@0.5), and the frame-level overlap decision by accuracy, precision, recall, and F1, with overlap as the positive class. We refer to mAP@0.5 and overlap F1 together as detection quality.

\subsection{YOLO Detectors and Training}

Modern YOLO detectors consist of a backbone that downsamples the input, a neck that fuses features across resolutions, and a head that predicts boxes and classes at multiple scales~\cite{ramos2025yolo} (Fig.~\ref{fig:yolo_mdi_vs_local}(a)). We evaluate YOLOv8, YOLO11, and YOLO26~\cite{ramos2025yolo,jocher2026yolo26} at the nano (n), small (s), and extra-large (x) scales. These scales cover sensing tasks of different complexity: smaller models suit simple conditions and tight budgets, whereas larger models provide the capacity that more complex \cbrs{} environments, with more signal types, lower SINRs, or denser coexistence, may require. Each detector is fine-tuned from COCO-pretrained weights with Ultralytics~\cite{jocher2023ultralytics}, and \appx{B} describes the differences between generations and the training settings.

\subsection{Edge Testbed}
\label{sec:testbed}

\begin{figure}[!t]
\centering
\includegraphics[width=0.5\columnwidth]{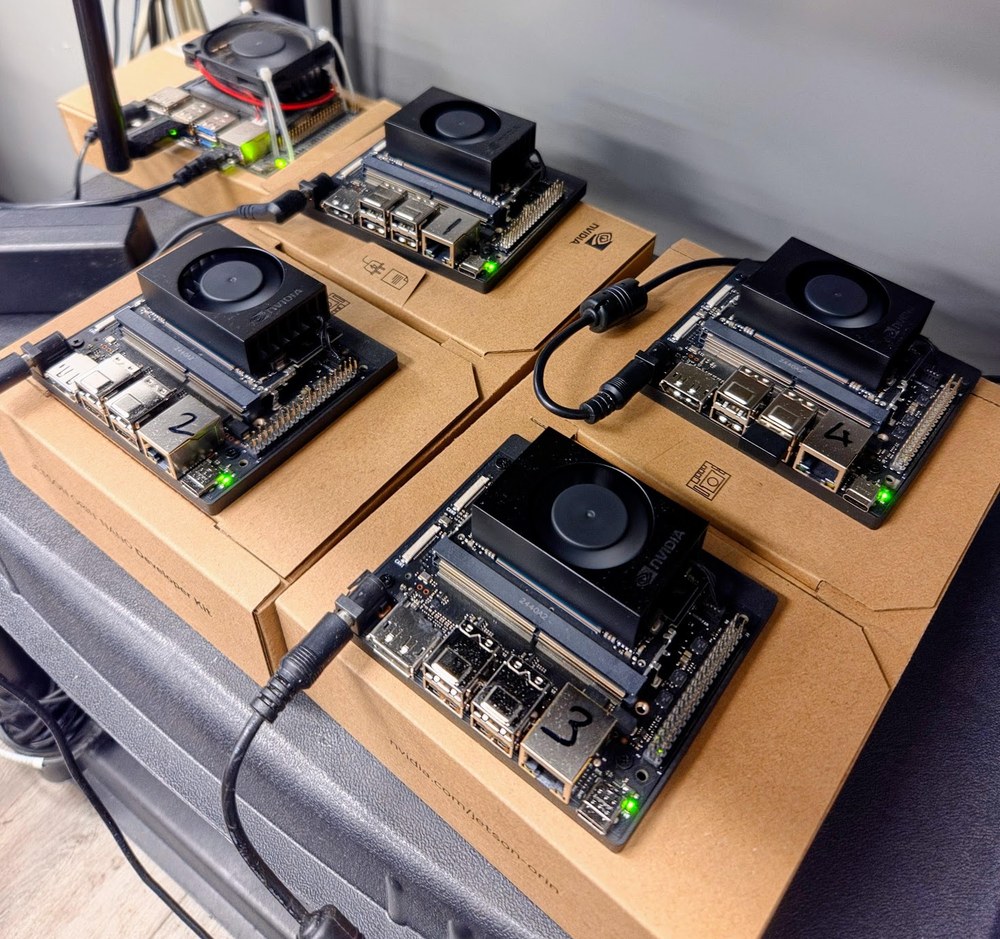}
\caption{Edge testbed with four NVIDIA Jetson Orin Nano Developer Kits.}
\label{fig:testbed}
\end{figure}

Our testbed (Fig.~\ref{fig:testbed}) consists of four NVIDIA Jetson Orin Nano Developer Kits connected through an 802.11ac wireless access point. The four devices serve as Nodes A--D of the distributed pipelines, and a single device runs the local-inference baselines. Although the Orin Nano includes a GPU, all experiments run on the CPU in FP32, because we target general edge devices, such as drones, routers, and surveillance cameras, many of which lack GPUs or other deep-learning accelerators. Application-layer token buckets create high-, medium-, and low-bandwidth environments with measured link rates of approximately 70, 38, and 21~Mbps (\appx{C}).

\subsection{Local Inference Performance}

Table~\ref{tab:yolo_cpu_results} shows that local inference achieves high detection quality: every detector reaches an mAP@0.5 and an overlap F1 of at least 0.991, and several are nearly perfect on every metric, whereas the earlier YOLOv3-based study~\cite{tassie2023cbrs} reported a radar recall of 86.6\%. Throughput, however, is only 0.24--0.88 frames/s on a single device, so we next accelerate the detectors with \mdi{}.

\begin{table}[!t]
\centering
\caption{Local inference results of the original YOLO models. Accuracy, precision, recall, and F1 refer to the frame-level overlap decision.}
\label{tab:yolo_cpu_results}
\footnotesize
\setlength{\tabcolsep}{2pt}
\renewcommand{\arraystretch}{0.8}
\resizebox{0.92\columnwidth}{!}{%
\begin{tabular*}{\columnwidth}{@{\extracolsep{\fill}}lrrrrrr@{}}
\toprule
Model & Thr. (frames/s) & Accuracy & Precision & Recall & F1 & mAP@0.5 \\
\midrule
YOLOv8n & 0.846 & 0.996 & 1.000 & 0.995 & 0.997 & 0.995 \\
YOLOv8s & 0.671 & 0.999 & 1.000 & 0.998 & 0.999 & 0.995 \\
YOLOv8x & 0.243 & 0.993 & 1.000 & 0.989 & 0.995 & 0.995 \\
YOLO11n & 0.875 & 0.996 & 1.000 & 0.995 & 0.997 & 0.995 \\
YOLO11s & 0.688 & 1.000 & 1.000 & 1.000 & 1.000 & 0.995 \\
YOLO11x & 0.260 & 0.993 & 1.000 & 0.989 & 0.995 & 0.995 \\
YOLO26n & 0.815 & 0.998 & 1.000 & 0.996 & 0.998 & 0.994 \\
YOLO26s & 0.628 & 0.993 & 1.000 & 0.989 & 0.995 & 0.995 \\
YOLO26x & 0.240 & 0.988 & 1.000 & 0.982 & 0.991 & 0.991 \\
\bottomrule
\end{tabular*}
}
\end{table}

\section{MDI of the Original YOLO}
\label{sec:yolo_mdi_exploration}

\subsection{Pipeline Model}
\label{sec:mdi_pipeline_model}

We model \mdi{} as a pipeline over a chain of nodes. The first node is the sensing endpoint, which captures and preprocesses the I/Q samples. Each subsequent node executes a consecutive part of the detector, and the last node returns the detections to the sensing endpoint. The model is not tied to a particular number of nodes or hardware platform. Our testbed uses Nodes A--D, with Node A as the sensing endpoint. The runtime exchanges data over HTTP and buffers frames in first-in, first-out queues, so the computation and communication of different frames overlap. All distributed experiments use this runtime (\appx{C}).

We measure throughput as \(N/T_{\mathrm{run}}\) for \(N\) frames completed in run time \(T_{\mathrm{run}}\), including pipeline filling and draining. Our figures report its inverse, the inference time \(T_{\mathrm{run}}/N\), so a shorter inference time means a higher throughput. The inference time characterizes the inference system, whereas the delay of a single frame is its end-to-end delay defined below. Let \(t_i\) denote the compute time of stage \(i\), \(b_i\) the activation payload transferred after it, and \(B_i\) the corresponding link rate. To first order, a filled pipeline achieves the throughput
\begin{equation}
R_{\mathrm{pipe}} \approx \left(\max_i \{t_i, b_i/B_i\}\right)^{-1},
\label{eq:mdi_pipeline_rate}
\end{equation}
so throughput is governed by the slowest computation or transfer. Pipelining raises throughput at the cost of queueing delay, which adds to the end-to-end delay of each frame, the time from the input of a frame to the output of its detections, including computation, communication, and queueing delays.

\subsection{Module-Boundary Partitioning}

Existing \mdi{} methods, such as~\cite{li2025pamdi}, treat the model as a chain of layers, so a cut between any two layers yields consecutive partitions that pass data only forward to the next node. The original YOLO is not a chain: its neck repeatedly fuses the backbone features P3, P4, and P5 across scales, so any cut after P3 is crossed by several branches. We therefore partition the detectors at module boundaries: Node B executes the backbone, Node C the neck, and Node D the head, so each inner cut point carries three feature maps, P3--P5 or N3--N5 (Fig.~\ref{fig:yolo_mdi_vs_local}(a)).

Without compression, this partition preserves the local overlap F1 of every detector, but in the low-bandwidth environment its throughput is only \(0.40\text{--}0.75\times\) that of local inference (Fig.~\ref{fig:yolo_mdi_vs_local}(c)). Four devices are thus slower than one, because communication rather than computation limits the pipeline.

\subsection{Activation Compression}
\label{sec:activation_compression}

Compressing the representations exchanged between workers reduces the communication cost of \mdi{}~\cite{communication_aware_inference_2026}. We apply lightweight compressors directly to the exchanged activations: top-k sparsification with 25\%, 50\%, or 75\% retention, INT8 and INT4 quantization~\cite{communication_aware_inference_2026}, and a hybrid compressor that follows LLM.int8()~\cite{dettmers2022llmint8} by keeping outliers in FP16 and quantizing the remaining values to INT8 (\appx{D}).

\begin{figure*}[!t]
\centering
\includegraphics[width=\textwidth]{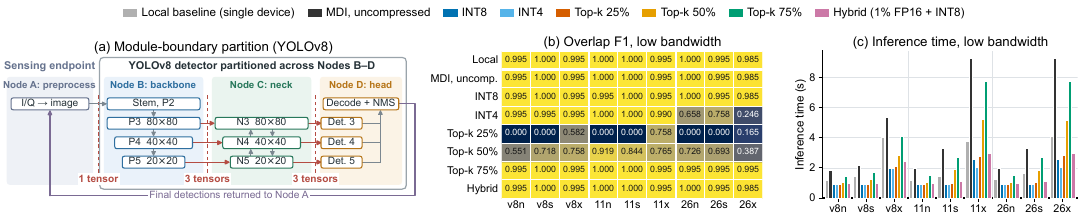}
\caption{MDI of the original YOLO compared with local inference (\appx{D}). (a) Module-boundary partition of YOLOv8, with the number of tensors crossing each cut point. (b) Overlap F1 and (c) inference time (inverse of throughput) in the low-bandwidth environment.}
\label{fig:yolo_mdi_vs_local}
\end{figure*}

Fig.~\ref{fig:yolo_mdi_vs_local} shows that suitable compression substantially reduces the communication cost with little effect on detection quality. INT8 reduces the payload to 25\% and raises throughput over uncompressed \mdi{} by up to 264\% while changing overlap F1 by at most 0.005, so every detector exceeds local throughput without losing detection quality. The hybrid compressor matches the uncompressed F1 for every detector. More aggressive compression degrades quality: top-k with 25\% or 50\% retention lowers F1 for all detectors, and INT4 lowers the F1 of YOLO26 to 0.25--0.76. Top-k with 75\% retention preserves quality but still transmits 78\% of the uncompressed payload.

\subsection{Structural Limitations}
\label{sec:yolo_mdi_limits}

Two structural properties of the original YOLO limit \mdi{} regardless of the compressor. First, every inner cut point is crossed by three multi-scale feature maps, so the transferred volume is large and grows with model scale. Second, the cut points are fixed at module boundaries rather than chosen by cost. For the extra-large detectors, the backbone node carries 40--46\% of the computation, 1.7--1.9~s per frame, so by~\eqref{eq:mdi_pipeline_rate} even unlimited bandwidth caps throughput at \(2.0\text{--}2.4\times\) that of local inference, far below the ideal of \(4\times\) with four balanced nodes.

The original YOLO is thus poorly suited to \mdi{}: its multi-scale graph restricts the cut points, each cut point carries several large feature maps, and the resulting partitions overload a single node. Compression reduces the communication cost but can lower detection quality, and it cannot fix the load imbalance. A good partition of the original YOLO also depends on the hardware and the network topology, which motivates \cedaryolo{}, a detector with a more flexible partitioning structure (Section~\ref{sec:cedar_yolo}). \system{} then exploits its structure to balance the pipeline (Section~\ref{sec:cedar_yolo_mdi}).

\section{CEDAR-YOLO: A Single-Scale RF-Aware Detector}
\label{sec:cedar_yolo}

\cedaryolo{} applies structured pruning at the architecture level: instead of removing individual weights, it removes entire modules of YOLOv8, namely the multi-scale neck and two of the three detection scales. \cbrs{} signals occupy a small set of known bandwidths, so their boxes vary far less in scale than objects in natural images, and a single prediction scale suffices. \cedaryolo{} then adapts the remaining backbone and detection head to the separate frequency and time axes of radar/LTE spectrograms. Because no pretrained weights exist for the pruned architecture, we train it from scratch (\appx{E}).

\subsection{Architecture}

\begin{figure}[!t]
\centering
\includegraphics[width=\columnwidth]{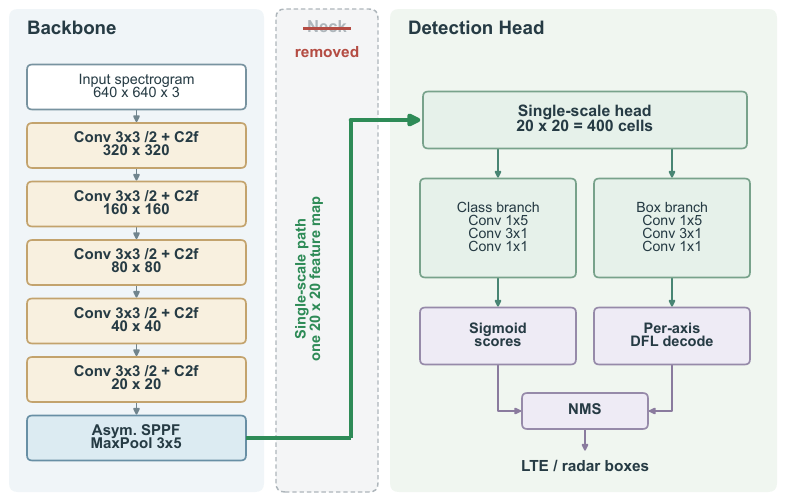}
\caption{\cedaryolo{} architecture with a \(20\times20\) grid. Compared with Fig.~\ref{fig:yolo_mdi_vs_local}(a), the multi-scale neck and two of the three detection scales are removed. Other grids change the backbone stages.}
\label{fig:cedar_yolo_architecture}
\end{figure}

As shown in Fig.~\ref{fig:cedar_yolo_architecture}, \cedaryolo{} is a single chain of backbone stages, each consisting of a strided convolution and a C2f block, followed by asymmetric spatial pyramid pooling and one detection head. The head predicts boxes on an \(F\times T\) output grid with \(F\) frequency cells and \(T\) time cells, and regresses the frequency and time extents of each box separately. Whereas every YOLO stage halves both axes, a \cedaryolo{} stage halves either both axes or only one, so \(F\) and \(T\) can be chosen independently. The stages first halve both axes until the axis with more cells reaches its target and then halve only the other axis, so the axis with fewer cells sets the number of stages. For a \(640\times640\) input, a \(20\times20\) grid uses five stages that halve both axes and a \(40\times40\) grid uses four, whereas a \(20\times80\) grid uses three such stages followed by two that halve only the frequency axis (\appx{E}). The grid therefore sets both the resolution along each axis and the depth of the backbone. Because the network is a single chain, every block boundary carries exactly one feature tensor, which \system{} exploits for flexible partitioning (Section~\ref{sec:cedar_yolo_mdi}).

\subsection{Grid and Coverage Selection}

\begin{figure*}[!t]
\centering
\includegraphics[width=\textwidth]{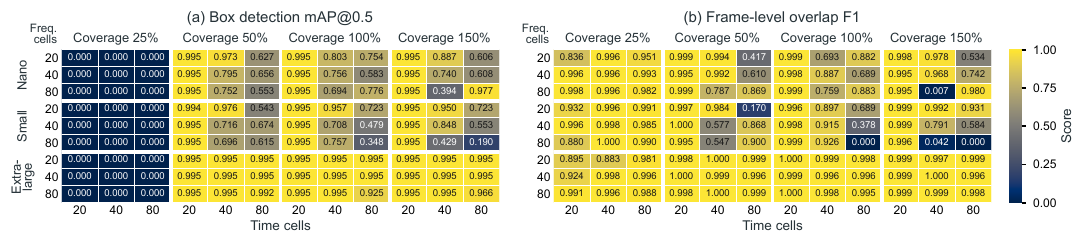}
\caption{\cedaryolo{} parameter sweep on the held-out test set. (a) Box detection mAP@0.5. (b) Frame-level overlap F1. Rows group the model scale and the number of frequency cells, and columns group the regression coverage and the number of time cells.}
\label{fig:cedar_yolo_grid_sweep}
\end{figure*}

Radar and LTE signals differ in bandwidth and duration, so the resolution that best captures their features along each axis, and the regression range needed to span their boxes, are not known in advance. We therefore sweep 108 configurations: three model scales, 20, 40, or 80 cells along each axis, and regression coverages of 25\%, 50\%, 100\%, and 150\%. Coverage sets the number of regression bins, and thus the largest distance from a cell to a box side, as a fraction of the number of cells along each axis. Fig.~\ref{fig:cedar_yolo_grid_sweep} reports mAP@0.5 and overlap F1, and \appx{E} adds accuracy, precision, recall, and the size of each configuration. Once the coverage reaches 50\%, most configurations detect the signals well: 52 of these 81 configurations reach an overlap F1 of at least 0.95, and the extra-large model reaches an mAP@0.5 of at least 0.925 and an overlap F1 of at least 0.994 with every grid. We use the \(20\times20\) grid in the remaining experiments, the natural default since it matches the stride-32 P5 output of YOLOv8. It is also among the best, with an mAP@0.5 of 0.994--0.995 and an overlap F1 of at least 0.996 at every scale.

\subsection{Comparison with YOLOv8}

\begin{figure}[!t]
\centering
\includegraphics[width=\columnwidth]{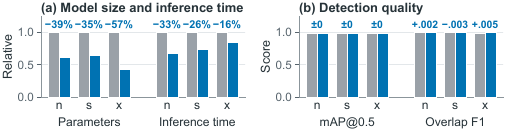}
\caption{\cedaryolo{} (\(20\times20\) grid, 100\% coverage, blue) compared with YOLOv8 (gray) at each scale. (a) Parameter count and model inference time relative to YOLOv8, with reductions labeled. (b) mAP@0.5 and overlap F1, with changes labeled.}
\label{fig:cedar_yolo_local_comparison}
\end{figure}

Fig.~\ref{fig:cedar_yolo_local_comparison} compares the \(20\times20\) \cedaryolo{} at 100\% regression coverage with the original YOLOv8 of the same scale. \cedaryolo{} reduces the parameter count by 39\%, 35\%, and 57\% at the nano, small, and extra-large scales, respectively, and the model inference time by 33\%, 26\%, and 16\%. Its mAP@0.5 equals that of YOLOv8 at every scale, and its overlap F1 differs by less than 0.006.

\section{CEDAR-MDI: Compression-Aware Model-Distributed Inference}
\label{sec:cedar_yolo_mdi}

\begin{figure*}[!t]
\centering
\includegraphics[width=\textwidth]{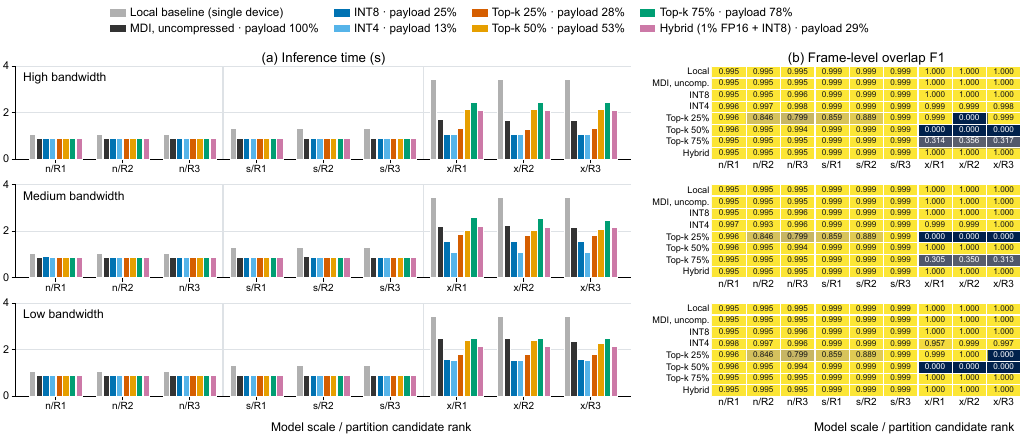}
\caption{\system{} with candidates R1--R3 of each setting compared with local inference on the full test set, one row per bandwidth environment. A tick such as x/R2 denotes the extra-large model with its second-ranked candidate. For n and s, the preprocessing on Node A is the bottleneck. (a) Inference time, the inverse of throughput. (b) Frame-level overlap F1. The legend gives the median payload ratio used by the optimizer.}
\label{fig:cedar_yolo_mdi_comparison}
\end{figure*}

\subsection{Partitionable Execution Path}

\system{} runs the nano, small, and extra-large \cedaryolo{} models with the \(20\times20\) grid and 100\% regression coverage on the runtime of Section~\ref{sec:mdi_pipeline_model}. The profiler splits each model, from raw I/Q preprocessing to post-processing, into 16 ordered stages, and a partition assigns nonempty, contiguous groups of stages to the nodes in route order, Nodes A--D on our testbed. Unlike in Section~\ref{sec:yolo_mdi_exploration}, where Node A runs all preprocessing, \system{} can also distribute the preprocessing, which takes a large share of the local inference time, especially for the smaller models (Fig.~\ifdefined\uselong\ref{fig:yolo_cpu_stage_cost}\else B.2\fi{} in \appx{B}). Every boundary carries a single tensor, and the compressors of Section~\ref{sec:activation_compression} are applied to the input tensor and the intermediate features (\appx{F}).

\subsection{Cost Profiling and Partition Optimization}
\label{sec:cedar_offline_policy}

We profile the execution time and output size of every stage on a Jetson, the effective rate of each link in each bandwidth environment, and the compression ratio of each compressor (\appx{F}). A policy \(p\) partitions the ordered stages into contiguous groups and assigns them to the nodes \(v\in\mathcal{V}\) in route order. Let \(\mathcal{S}_v(p)\) denote the stages of node \(v\), and \(\mathcal{L}\) the forward links and the return link. Given the stage compute times \(t_s\), the uncompressed boundary sizes \(b_\ell(p)\), the compression ratios \(r_\ell(p,q)\) of compressor \(q\), and the link rates \(B_\ell\), we select the policy for each compressor \(q\) by solving
\begin{equation}
\begin{array}{ll}
\displaystyle\min_{p\in\mathcal{P},\,\tau} & \tau \\[2pt]
\mathrm{subject\ to}
 & \displaystyle\sum_{s\in\mathcal{S}_v(p)}t_s\leq\tau,
   \quad v\in\mathcal{V}, \\[4pt]
 & \displaystyle\frac{b_\ell(p)r_\ell(p,q)}{B_\ell}\leq\tau,
   \quad \ell\in\mathcal{L},
\end{array}
\label{eq:cedar_partition_optimization}
\end{equation}
where \(\mathcal{P}\) is the feasible set and \(\tau\) is the estimated pipeline service time. Minimizing \(\tau\) maximizes the predicted throughput \(1/\tau\) of~\eqref{eq:mdi_pipeline_rate}. With 16 stages and four nodes, \(\mathcal{P}\) contains at most \(\binom{15}{3}=455\) policies, so we evaluate every feasible policy offline and rank the policies by \(\tau\). Because the profiled costs are estimates, the measured ranking can differ from the predicted one, so we deploy the three highest-ranked candidates, R1--R3, for each model, bandwidth environment, and compressor.

The optimization targets throughput rather than end-to-end delay, which only has to stay well within the time allowed for \cbrs{} sensing (Section~\ref{sec:cedar_distributed}).

\subsection{Distributed Performance}
\label{sec:cedar_distributed}

Fig.~\ref{fig:cedar_yolo_mdi_comparison} shows all three deployed candidates of each setting on the full test set.

\emph{Detection quality.} Uncompressed \mdi{}, INT8, INT4, and hybrid compression preserve the local overlap F1 in nearly all settings, whereas top-k sparsification is unstable across boundaries and can lose much of the detection quality (Fig.~\ref{fig:cedar_yolo_mdi_comparison}(b)), so we do not recommend it. Below, we consider only runs whose overlap F1 is within 0.01 of local inference.

\emph{Throughput.} These runs exceed local throughput for every model, bandwidth environment, compressor, and candidate rank. For the extra-large model, the top-ranked candidates reach up to \(3.21\times\) the local throughput, and the best of the three candidates reaches \(3.27\times\), with the same overlap F1 as local inference. At this scale, \system{} reaches 0.95 frames/s, compared with at most 0.53 frames/s for \mdi{} of the original YOLO (\appx{D}). For the nano and small models, the preprocessing that remains on Node A, about 0.89~s per frame, is the bottleneck, since every other node and link takes less time. Preprocessing varies widely across receivers and data formats and is outside our scope, which centers on the detector. Two properties of \cedaryolo{} explain the gains over \mdi{} of the original YOLO.

\emph{Partition flexibility.} The original YOLO admits a single module-boundary partition, whereas the single-scale chain of \cedaryolo{} offers many candidate partitions, each crossing only one tensor per boundary. The optimizer can therefore place cut points within the preprocessing and the early backbone. Without compression, the selected partitions transfer 2.5--2.9, 3.3--3.7, and 10.0--19.1~MB per frame for the nano, small, and extra-large models, compared with 10.7--12.3, 16.4--19.7, and 31.5--46.7~MB for \mdi{} of the original YOLO.

\emph{Load balance.} For the extra-large model in the high-bandwidth environment with INT8, the selected partition assigns 0.71--1.04~s of computation per frame to each node, so the busiest node carries 29\% of the total and the nodes are busy 85\% of the time on average. \mdi{} of the original extra-large detectors instead leaves 1.7--1.9~s on the backbone node, 40--46\% of the total, and keeps the nodes busy at most 63\% of the time.

\emph{End-to-end delay.} Pipelining adds queueing delay, but the delay stays well within the time allowed for \cbrs{} sensing: even the largest end-to-end delay, 16.1~s overall and 7.1~s at the highest throughput, is far below the 60~s that WInnForum allows for radar detection~\cite{winnf2017cbrsreq} and the 300~s within which the FCC requires the affected \cbrs{} devices to be suspended or relocated~\cite{fcc2026part96}.

\section{Conclusion}
\label{sec:conclusion}

We presented CEDAR, a communication-efficient, edge-distributed architecture for YOLO-based \cbrs{} radar detection. We first built an end-to-end pipeline from raw I/Q samples to frame-level overlap decisions and showed that recent YOLO generations achieve an overlap F1 above 0.99 on real-world \cbrs{} data. Through analysis and testbed experiments, we then showed that the original YOLO is poorly suited to \mdi{}: its multi-scale structure restricts the cut points, inflates the transferred data, and overloads a single node. \cedaryolo{} prunes this structure into a single-scale detector with 35--57\% fewer parameters without sacrificing detection quality, and because every block boundary carries a single tensor, it enables flexible partitioning. \system{} compresses the exchanged activations and, for each compressor, selects offline the partition that minimizes the profiled pipeline bottleneck. On a testbed of NVIDIA Jetson Orin Nano devices, \system{} balances the computation across nodes, reduces the transferred data by up to 80\% compared with \mdi{} of the original YOLOv8, and achieves up to \(3.27\times\) the throughput of local inference while preserving detection quality, with end-to-end delays far below the \cbrs{} timing requirements.

\balance
\bibliographystyle{IEEEtran}
\bibliography{refs}

\clearpage
\newgeometry{margin=1in,twocolumn=false}
\onecolumn
\appendices
\numberwithin{figure}{section}
\numberwithin{table}{section}
\numberwithin{equation}{section}

\section{Dataset, Preprocessing, and Evaluation}
\label{app:A}

The real-world OTA \cbrs{} dataset of~\cite{tassie2023cbrs} contains controlled overlapping and non-overlapping radar/LTE scenarios stored as paired SigMF data and metadata files with annotated LTE and radar boxes. Our pipeline turns each frame into a spectrogram image, detects and localizes the LTE and radar signals with class-labeled boxes in time-frequency space, and decides from the boxes whether the two signals overlap in frequency (Fig.~\ref{fig:cbrs_detection_examples}). Table~\ref{tab:app_dataset} summarizes the dataset and our split. We split the frames within each LTE bandwidth so that every set contains all four LTE bandwidths.

Our experiments focus on this open dataset, which provides real OTA captures with annotated boxes and allows a reproducible comparison of detectors and distributed configurations. Its scenarios are controlled, with SINRs of 15--35~dB and one LTE and one radar signal per frame. In future work, we will evaluate the detectors on real-time data captured with software-defined radio (SDR) hardware, which also allows lower SINRs, frames without radar, and additional interference sources.

\begin{table}[!ht]
\centering
\caption{Dataset and split.}
\label{tab:app_dataset}
\begin{tabular}{ll}
\toprule
Item & Value \\
\midrule
Frames & 5,640 SigMF data/metadata pairs \\
Frame content & 40~ms of complex I/Q samples at 30.72~MS/s, centered at 3.6~GHz \\
LTE bandwidth & 5, 10, 15, or 20~MHz \\
Radar bandwidth & 2~MHz \\
SINR & 15--35~dB \\
Split & 70/15/15 within each LTE bandwidth, fixed random seed \\
Training / validation / test frames & 3,946 / 846 / 848 \\
Test frames, overlapping / non-overlapping & 558 / 290 \\
\bottomrule
\end{tabular}
\end{table}

Preprocessing converts each frame into a \(640\times640\) log-magnitude spectrogram image and then into the model input in the seven steps of Table~\ref{tab:app_preprocessing}. In the image, the horizontal coordinate of a box encodes its occupied frequency interval within the 30.72-MHz observation bandwidth, and the vertical coordinate encodes its time range within the 40-ms frame. Preprocessing is outside the scope of this work, which centers on the detector. It depends strongly on the receiver hardware and the data format, such as the sampling rate, the I/Q representation, and whether the spectrogram is rendered as an image, so we use a straightforward implementation and do not optimize it.

\begin{table}[!ht]
\centering
\caption{Preprocessing steps from raw I/Q samples to the model input.}
\label{tab:app_preprocessing}
\begin{tabular}{cll}
\toprule
Step & Operation & Output \\
\midrule
1 & Read the complex I/Q samples & Complex I/Q vector \\
2 & 512-point spectrogram with 384-sample overlap (SciPy) & Magnitude spectrogram \\
3 & Frequency shift and logarithmic conversion (NumPy) & Log-magnitude spectrogram \\
4 & 5th--95th percentile rendering limits & Spectrogram with rendering limits \\
5 & Render the image (Matplotlib) & RGBA canvas \\
6 & Convert the canvas to RGB (Pillow) & RGB image \\
7 & Build the floating-point input tensor & \(1\times3\times640\times640\) tensor \\
\bottomrule
\end{tabular}
\end{table}

We evaluate all detectors on the held-out test set. Box detection is measured by mAP@0.5, the mean average precision over the LTE and radar classes at an IoU threshold of 0.5. The overlap decision is evaluated per frame with overlap as the positive class. Accuracy is the fraction of frames whose overlap state is predicted correctly, precision is the fraction of predicted overlapping frames that truly overlap, recall is the fraction of overlapping frames that are predicted as overlapping, and F1 is the harmonic mean of precision and recall. The horizontal extents of the LTE and radar boxes define their occupied frequency intervals, and a frame is overlapping if any LTE--radar interval pair has a positive-width intersection. The same rule is applied to the ground-truth and predicted boxes, so the predicted overlap state can be compared directly with the reference state.

\FloatBarrier
\section{Original YOLO Models and Training}
\label{app:B}

\begin{figure}[!ht]
\centering
\includegraphics[width=0.62\textwidth]{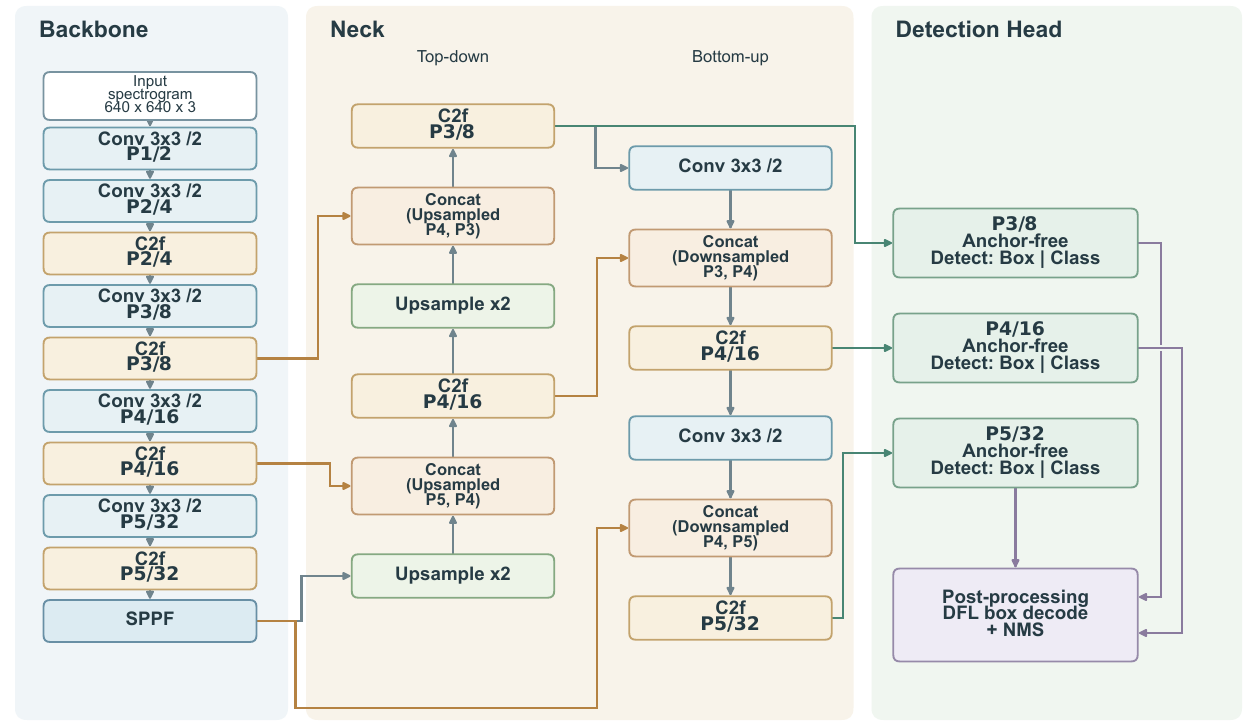}
\caption{Simplified schematic of the YOLOv8 architecture.}
\label{fig:yolov8_architecture}
\end{figure}

YOLO formulates object detection as a one-stage prediction problem: a single network maps an input image directly to class scores and bounding boxes without a separate region-proposal stage. Modern YOLO detectors organize this computation into a backbone, a neck, and a detection head (Fig.~\ref{fig:yolov8_architecture}). The backbone progressively downsamples the input to extract hierarchical features, the neck combines features across resolutions through top-down and bottom-up paths, and the head predicts boxes and classes at multiple scales~\cite{ramos2025yolo}. Within each generation, the \texttt{n}, \texttt{s}, and \texttt{x} variants share the same design and differ in depth, channel width, and computational cost. Table~\ref{tab:app_yolo_generations} summarizes how the three generations differ: YOLO11 refines the feature blocks of YOLOv8, and YOLO26 additionally changes the prediction interface so that inference no longer needs NMS.

\begin{table}[!ht]
\centering
\caption{Differences among the evaluated YOLO generations~\cite{ramos2025yolo,jocher2026yolo26}.}
\label{tab:app_yolo_generations}
\begin{tabular}{llll}
\toprule
 & YOLOv8 & YOLO11 & YOLO26 \\
\midrule
Feature block & C2f & C3k2 & C3k2, attention in the final P5 neck block \\
Backbone tail & SPPF & SPPF and C2PSA & As YOLO11, with a shortcut in SPPF \\
Neck & FPN--PAN & FPN--PAN & FPN--PAN \\
Box regression & DFL & DFL & Direct regression \\
Training heads & One-to-many & One-to-many & One-to-many and one-to-one \\
Inference output & Dense, with NMS & Dense, with NMS & One-to-one with top-k, without NMS \\
\bottomrule
\end{tabular}
\end{table}

We fine-tune each original detector from its COCO-pretrained Ultralytics weights on the fixed split with the settings of Table~\ref{tab:app_yolo_training}. Ultralytics~\cite{jocher2023ultralytics} is also used for their local and distributed inference. The architectures are unchanged during fine-tuning, and the checkpoint with the best validation performance is retained for evaluation.

\begin{table}[!ht]
\centering
\caption{Training settings of the original YOLO detectors.}
\label{tab:app_yolo_training}
\begin{tabular}{ll}
\toprule
Setting & Value \\
\midrule
Initialization & COCO-pretrained Ultralytics weights \\
Input size & \(640\times640\) \\
Epochs / batch size & 25 / 16 \\
Optimizer & AdamW, \(\beta_1=0.9\), \(\beta_2=0.999\), weight decay \(5\times10^{-4}\) \\
Learning rate & \(1.667\times10^{-3}\), linear decay to 1\% of the base rate \\
Warmup & 3 epochs from zero, with \(\beta_1\) rising from 0.8 to 0.9 \\
Precision and reproducibility & Automatic mixed precision, seed 0, deterministic algorithms \\
Color augmentation & Hue 0.015, saturation 0.7, value 0.4 \\
Geometric augmentation & Translation 10\%, scale 0.5--1.5, horizontal flip with probability 0.5 \\
Mosaic & Probability 1.0, disabled for the final 10 epochs \\
Disabled augmentation & Rotation, shear, perspective, vertical flip, MixUp, CutMix, Copy-Paste \\
\bottomrule
\end{tabular}
\end{table}

\begin{figure}[!ht]
\centering
\includegraphics[width=0.62\textwidth]{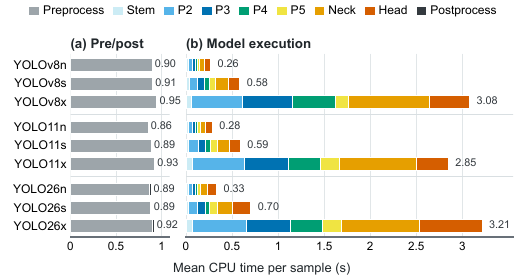}
\caption{Mean CPU time per frame of each stage of the YOLO detector pipeline, which together make up the local inference time. (a) Preprocessing and post-processing. (b) Model computation, divided into the stem, P2--P5 backbone stages, neck, and detection head.}
\label{fig:yolo_cpu_stage_cost}
\end{figure}

Figure~\ref{fig:yolo_cpu_stage_cost} breaks down the local inference time. Preprocessing and post-processing take a similar time for every detector, whereas the model computation grows with scale. The backbone accounts for about half of the model inference time, and its cost grows fastest with scale, so most of the additional cost of larger variants arises before multi-scale fusion and prediction. A direct backbone--neck--head partition therefore assigns very different loads to the participating devices.

\FloatBarrier
\section{Testbed and Distributed Runtime}
\label{app:C}

The testbed consists of four NVIDIA Jetson Orin Nano Developer Kits, which act as Nodes A--D and exchange data over HTTP through an 802.11ac wireless access point. Each device has a six-core Arm Cortex-A78 CPU, an Ampere-class integrated GPU, and 8~GB of 128-bit LPDDR5 memory. All experiments, including the local baselines and every node of the distributed pipelines, execute on the CPU in FP32, and the GPU is not used. CPU support for FP16 and other reduced-precision arithmetic is limited on general edge devices, and FP32 is also the default precision of Ultralytics inference.

We create the three bandwidth environments with application-layer token buckets. Tokens accumulate at the configured rate, and each serialized HTTP message consumes tokens according to its size and waits when the balance is insufficient. The achieved rate also depends on the wireless channel and HTTP processing. In our preliminary runs, the median rates of the three forward links are 64.6--75.1, 36.5--38.8, and 21.1--21.5~Mbps in the high-, medium-, and low-bandwidth environments, which average to approximately 70, 38, and 21~Mbps.

\emph{Runtime.} Every node buffers frames in first-in, first-out queues, so the computation and communication of different frames overlap. Node A admits a new frame, runs its assigned stages, and places the result in a send queue. Every subsequent node runs three workers: a receive worker appends arriving frames to an inference queue, an inference worker processes one frame at a time and places its output in a send queue, and a send worker transmits one frame at a time to the next node. The last node returns the detections to Node A, where a separate quality worker compares them with the ground truth.

\emph{In-flight limit.} Node A admits a new frame only when one of \(K\) in-flight slots is free and releases the slot when the detections of that frame return, so at most \(K\) frames are in the system and every queue holds at most \(K\) frames. The limit prevents the queues from growing too long. Without it, Node A would keep admitting frames faster than the slowest stage processes them, and the queue in front of that stage would keep growing, holding more and more frames and their intermediate tensors in the limited memory of the edge devices. The limit keeps enough frames in flight for computation and communication to overlap while bounding the queue length and the memory and other resources each node spends on waiting frames. All experiments use \(K=6\).

\emph{Throughput and delay.} Throughput \(N/T_{\mathrm{run}}\) is measured from the start of a run to the return of its final frame. For sequential local execution, it equals the inverse of the mean per-frame time. Its inverse, the inference time \(T_{\mathrm{run}}/N\), reflects how often frames are completed rather than how long a single frame takes. Once multiple frames are in flight, throughput is governed by the slowest stage, and frames wait in the queue in front of it. The end-to-end delay \(D\) of a frame, from its admission at Node A to the return of its detections, therefore contains both the computation and transfer times along the path and the queueing delay \(w_i\) before each stage \(i\):
\begin{equation}
D = \sum_i \left(t_i + \frac{b_i}{B_i}\right) + \sum_i w_i.
\label{eq:mdi_end_to_end_delay}
\end{equation}

\FloatBarrier
\section{MDI of the Original YOLO}
\label{app:D}

We use \mdi{} rather than data-distributed inference because each input is large, 9.8~MB of raw I/Q samples or a 4.9-MB input tensor per frame, so sending every frame to a worker that runs the whole model is costly at the edge, as shown in~\cite{li2023armdi}.

\emph{Partitioning.} Existing \mdi{} formulations, including~\cite{li2025pamdi}, treat a network as a chain of layers: each worker receives the output of the preceding partition, executes a consecutive block of layers, and forwards its output to the next worker. YOLO is instead a directed acyclic graph with extensive cross-scale feature reuse (Fig.~\ref{fig:yolov8_architecture}), so a partition boundary must carry every feature map that downstream branches need. We therefore cut at the module boundaries of Table~\ref{tab:app_yolo_partition}. Node A sends the preprocessed image tensor to Node B, each inner cut point carries three feature maps, and Node D performs the detector-specific post-processing before returning the detections to Node A. Because the inner cut points carry multi-scale feature maps, the transferred volume is large and grows with model scale. Including the detection return, uncompressed \mdi{} transfers 10.7, 16.4, and 31.5~MB per frame for YOLOv8n, YOLOv8s, and YOLOv8x, and 12.3, 19.7, and 46.7~MB for the corresponding YOLO11 and YOLO26 models.

\begin{table}[!ht]
\centering
\caption{Module-boundary partition of the original YOLO detectors (layer indices of the Ultralytics graphs).}
\label{tab:app_yolo_partition}
\begin{tabular}{llll}
\toprule
 & & YOLOv8 & YOLO11, YOLO26 \\
\midrule
Node B & Backbone & Layers 0--9 & Layers 0--10 \\
Node C & Neck & Layers 10--21 & Layers 11--22 \\
Node D & Detection head & Layer 22 & Layer 23 \\
B\(\to\)C & P3, P4, P5 & Outputs of layers 4, 6, 9 & Outputs of layers 4, 6, 10 \\
C\(\to\)D & N3, N4, N5 & Outputs of layers 15, 18, 21 & Outputs of layers 16, 19, 22 \\
\bottomrule
\end{tabular}
\end{table}

\emph{Compressors.} Top-k sparsification keeps the given fraction of entries with the largest magnitudes and sends their positions in a PackBits-style bit mask~\cite{communication_aware_inference_2026}. Because the mask is also sent, retentions of 25\%, 50\%, and 75\% transmit 28\%, 53\%, and 78\% of the uncompressed payload. INT8 uses symmetric uniform quantization with a scale derived from the largest absolute value~\cite{communication_aware_inference_2026} and transmits 25\% of the payload. INT4 uses 4-bit values with stochastic rounding and transmits 13\%. Its stochastic rounding keeps the quantization error unbiased in expectation but causes small run-to-run fluctuations in detection quality. The hybrid compressor follows LLM.int8()~\cite{dettmers2022llmint8}: it stores the values above the 99th magnitude percentile in FP16 with a bit mask of their positions, quantizes the rest to INT8 with row-wise scales, and transmits about 30\%. Each compressor is applied to every forward transfer, including the A-to-B image tensor, and the receiving node reconstructs the tensors before executing its module.

\emph{Protocol.} The uncompressed runs process the full test set, and their overlap F1 and mAP@0.5 equal the local values for every detector and bandwidth environment. Because partitioning alone thus leaves the detection quality unchanged, the local and compressed runs process only 100 test frames, which suffice to measure the throughput. The compressors are therefore evaluated here only preliminarily. Their complete evaluation on the full test set follows later with \system{} (Section~\ref{sec:cedar_distributed} and Appendix~\ref{app:F}). For the uncompressed reference, overlap F1 uses the corresponding records of the full run, and throughput is that of the full run. Overlap F1 is recomputed from summed confusion counts.

\begin{figure}[!ht]
\centering
\includegraphics[width=\textwidth]{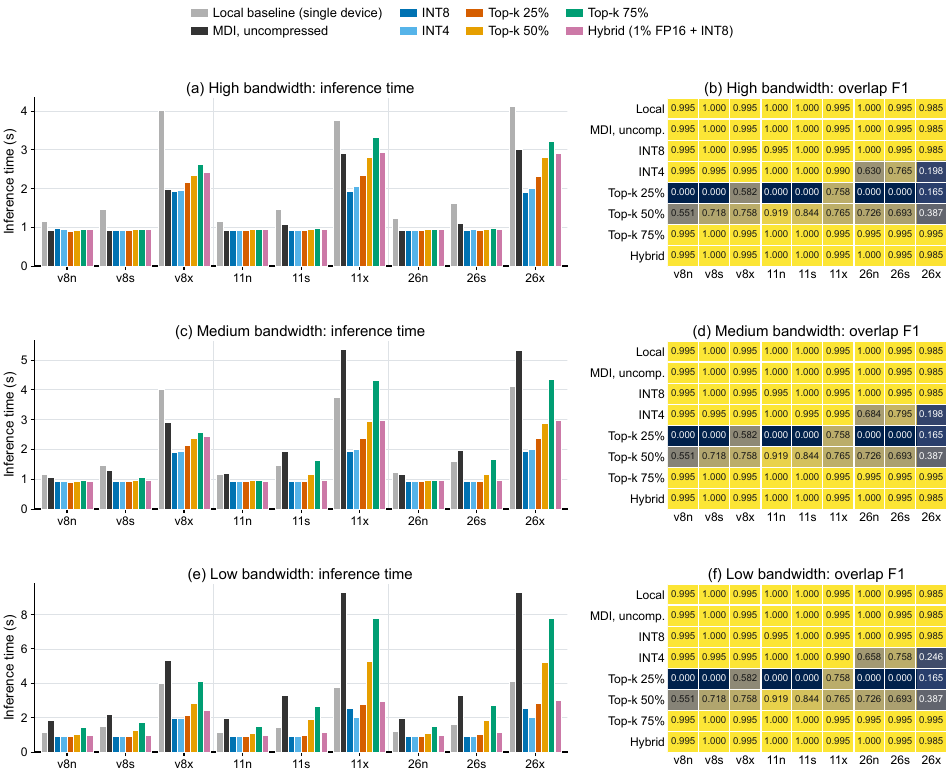}
\caption{MDI of the original YOLO compared with local execution on 100 test frames in all bandwidth environments. (a), (c), (e) Inference time, the inverse of throughput, in the high-, medium-, and low-bandwidth environments, where lower is better. (b), (d), (f) Frame-level overlap F1 of the same runs. Local denotes the single-device baseline, in which one Jetson executes the complete pipeline without partitioning, and is shown as the gray bar in each detector group.}
\label{fig:yolo_mdi_full}
\end{figure}

\emph{Results.} Figure~\ref{fig:yolo_mdi_full} covers all three bandwidth environments. Communication limits uncompressed \mdi{}, and more so as the bandwidth decreases: it reaches 1.26--2.02, 0.70--1.38, and 0.40--0.75 times the local throughput in the high-, medium-, and low-bandwidth environments. INT8 removes most of this limit. It lifts every detector above local throughput in every environment, to 1.19--2.17, 1.25--2.12, and 1.25--2.06 times, while changing overlap F1 by at most 0.005, and its maximum gain over uncompressed \mdi{} grows from 59\% to 177\% and 264\% as the bandwidth decreases. The hybrid compressor matches the uncompressed F1, whereas top-k with 25\% or 50\% retention lowers F1 for all detectors, and INT4 lowers the F1 of YOLO26 to 0.20--0.80. The F1 values are identical across environments except for INT4, which varies by at most 0.054 owing to its stochastic rounding. Even with compression, the extra-large detectors reach at most 0.53 frames/s, which is 94--96\% of the throughput cap set by their backbone node (Section~\ref{sec:yolo_mdi_limits}).

\FloatBarrier
\section{CEDAR-YOLO Details}
\label{app:E}

\emph{Architecture.} Starting from the YOLOv8 structure in Fig.~\ref{fig:yolov8_architecture}, we prune the top-down and bottom-up neck paths, including their upsampling, downsampling, and cross-scale concatenation, and replace the three prediction scales with a single head that follows the final backbone feature map and a spatial pyramid pooling block. Table~\ref{tab:app_cedar_components} specifies the remaining components. The pooling kernels and head convolutions are wider along frequency than along time, so they aggregate context across occupied bands, and the head regresses the frequency and time extents of each box separately.

\begin{table}[!ht]
\centering
\caption{\cedaryolo{} components.}
\label{tab:app_cedar_components}
\begin{tabular}{lp{0.68\textwidth}}
\toprule
Component & Configuration \\
\midrule
Backbone stage & \(3\times3\) convolution with stride (2,2), (2,1), or (1,2) along (frequency, time), followed by a C2f block, with batch normalization and SiLU \\
Asymmetric SPPF & \(1\times1\) convolution halving the channels, three successive \(3\times5\) (time \(\times\) frequency) max-pooling operations, concatenation, and a \(1\times1\) convolution back to the input channels \\
Detection head & Separate classification and box branches, each with \(1\times5\) frequency, \(3\times1\) time, and \(1\times1\) convolutions \\
Box regression & Left/right frequency and top/bottom time distances with per-axis distribution bins, decoded with the cell spacing of each axis \\
Post-processing & Confidence threshold 0.25 and class-wise NMS at IoU 0.45 \\
\bottomrule
\end{tabular}
\end{table}

\begin{figure}[!ht]
\centering
\includegraphics[width=\textwidth]{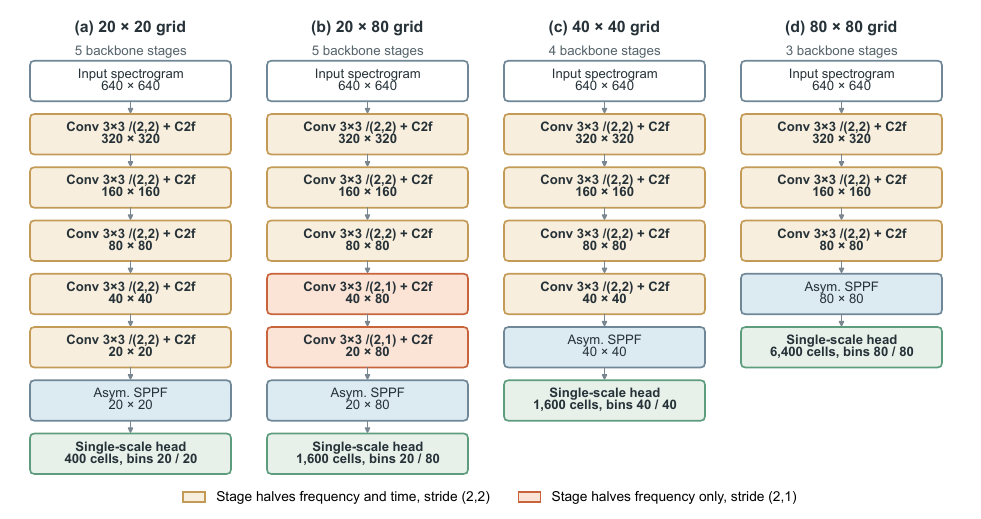}
\caption{\cedaryolo{} backbone layouts for four \(F\times T\) grids (frequency cells \(\times\) time cells). Each box gives the stage stride along (frequency, time) and the feature map size after the stage. The head uses one regression bin per cell on each axis (100\% coverage).}
\label{fig:cedar_grid_architectures}
\end{figure}

\emph{Grid configurations.} A stage halves both axes, only the frequency axis, or only the time axis. For a \(640\times640\) input and an \(F\times T\) grid, the frequency and time axes require \(n_F=\log_2(640/F)\) and \(n_T=\log_2(640/T)\) halvings. The backbone first applies \(\min(n_F,n_T)\) stages that halve both axes and then \(|n_F-n_T|\) stages that halve only the coarser axis, so it has \(\max(n_F,n_T)\) stages. Figure~\ref{fig:cedar_grid_architectures} shows the resulting layouts for four grids, and Table~\ref{tab:cedar_grid_configs} compares their size and quality. Refining both axes removes stages and thus parameters, but the larger feature maps keep the computation comparable or higher. Refining only one axis keeps the depth and the parameters, as in the \(20\times80\) grid, but the additional cells raise the computation. The \(20\times20\) grid gives the highest or tied-highest quality at every scale.

\begin{table}[!ht]
\centering
\caption{\cedaryolo{} configurations for four \(F\times T\) grids at 100\% regression coverage. Stages are given as the total and, in parentheses, the numbers that halve both axes and only the frequency axis. GFLOPs are counted for one \(640\times640\) input, and mAP@0.5 and overlap F1 come from the parameter sweep (Fig.~\ref{fig:cedar_sweep_full}).}
\label{tab:cedar_grid_configs}
\begin{tabular}{llcrrrrr}
\toprule
Grid (\(F\times T\)) & Scale & Stages & Final channels & Params (M) & GFLOPs & mAP@0.5 & Overlap F1 \\
\midrule
\(20\times20\) & n & 5 (5+0) & 256 & 1.85 & 4.0 & 0.995 & 0.999 \\
 & s & 5 (5+0) & 512 & 7.27 & 15.9 & 0.995 & 0.996 \\
 & x & 5 (5+0) & 640 & 29.17 & 159.0 & 0.995 & 1.000 \\
\midrule
\(20\times80\) & n & 5 (3+2) & 256 & 2.14 & 9.3 & 0.754 & 0.882 \\
 & s & 5 (3+2) & 512 & 7.55 & 34.0 & 0.723 & 0.689 \\
 & x & 5 (3+2) & 640 & 29.37 & 225.2 & 0.995 & 0.998 \\
\midrule
\(40\times40\) & n & 4 (4+0) & 128 & 0.66 & 4.0 & 0.756 & 0.887 \\
 & s & 4 (4+0) & 256 & 2.10 & 14.2 & 0.708 & 0.915 \\
 & x & 4 (4+0) & 640 & 18.53 & 162.3 & 0.995 & 0.996 \\
\midrule
\(80\times80\) & n & 3 (3+0) & 64 & 0.64 & 9.3 & 0.776 & 0.883 \\
 & s & 3 (3+0) & 128 & 1.14 & 18.6 & 0.348 & 0.000 \\
 & x & 3 (3+0) & 320 & 6.79 & 144.0 & 0.925 & 0.995 \\
\bottomrule
\end{tabular}
\end{table}

\emph{Training and evaluation.} \cedaryolo{} is trained and evaluated with our own PyTorch framework~\cite{cedarmdi2026code}, since its asymmetric backbone and single-scale head are not part of the Ultralytics model definitions. The \system{} runtime is likewise implemented independently. Table~\ref{tab:app_cedar_training} lists the training settings. The optimizer and loss weights follow the Ultralytics defaults, and the learning rate follows a cosine schedule. Unlike the fine-tuning of the original detectors, no data augmentation is applied. The averaged weights are validated after every epoch, and training stops early once the validation mAP@0.5 stops improving. For local timing, we run the \(20\times20\) models on raw SigMF inputs and measure the model inference time, i.e., the time of the forward pass. Because this time depends only on the architecture, the timings are paired with the quality of the corresponding sweep configuration.

\begin{table}[!ht]
\centering
\caption{Training settings of \cedaryolo{}.}
\label{tab:app_cedar_training}
\begin{tabular}{ll}
\toprule
Setting & Value \\
\midrule
Initialization & Random, since no pretrained weights exist for the pruned architecture \\
Data & Training and validation splits of Table~\ref{tab:app_dataset}, without data augmentation \\
Input size & \(640\times640\) \\
Epochs & Up to 100, with early stopping after 20 epochs without improvement in validation mAP@0.5 \\
Batch size & Depends on the model scale, with smaller batches for larger models \\
Optimizer & SGD with Nesterov momentum 0.937, weight decay \(5\times10^{-4}\) on convolution weights only \\
Learning rate & 0.01, cosine decay to 1\% of the base rate \\
Warmup & 3 epochs, linear \\
Gradient clipping & Maximum norm 10 \\
Precision & Automatic mixed precision \\
Weight averaging & Exponential moving average, decay 0.9999 with a 2,000-step ramp \\
Loss weights & Box overlap 7.5, binary cross-entropy classification 0.5, distribution focal 1.5 \\
Checkpoint selection & Best validation mAP@0.5 of the averaged weights \\
\bottomrule
\end{tabular}
\end{table}

\begin{figure}[!ht]
\centering
\includegraphics[width=\textwidth]{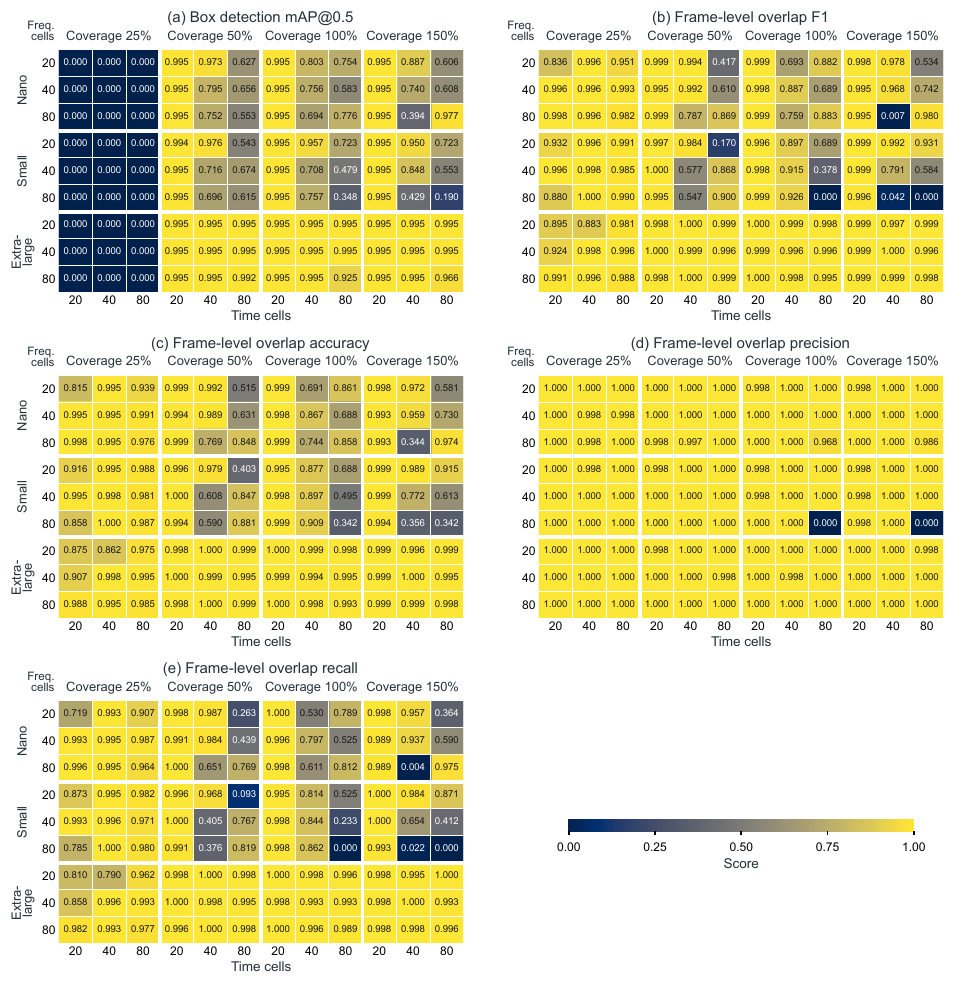}
\caption{Complete \cedaryolo{} parameter sweep on the held-out test set. (a) Box detection mAP@0.5. (b)--(e) Frame-level overlap F1, accuracy, precision, and recall. Rows group the model scale and the number of frequency cells, and columns group the regression coverage and the number of time cells. All panels share the same color scale.}
\label{fig:cedar_sweep_full}
\end{figure}

\emph{Parameter sweep.} The sweep combines the three model scales, 20, 40, or 80 frequency cells, 20, 40, or 80 time cells, and regression coverages of 25\%, 50\%, 100\%, and 150\%, which allocate 5, 10, 20, or 30 bins per box side at 20 cells, for 108 configurations in total. Figure~\ref{fig:cedar_sweep_full} reports all five quality metrics, with precision shown as 0 when no frame is predicted as overlapping. Coverage must be large enough for the bins to span the LTE and radar boxes: at 25\% coverage, mAP@0.5 is zero for every configuration, whereas the overlap decision, which depends only on frequency intervals, is less affected. At full coverage, coarse time grids keep the overlap F1 high across frequency resolutions. For the smaller models, swapping a fine-frequency, coarse-time grid for the reverse arrangement reduces the overlap quality, whereas the extra-large model is less sensitive. Because the grid also sets the backbone depth (Table~\ref{tab:cedar_grid_configs}), it controls both the spatial resolution and the model capacity.

\FloatBarrier
\section{CEDAR-MDI Details}
\label{app:F}

\emph{Execution path.} The profiler represents each deployed model as the 16 ordered stages of Table~\ref{tab:app_cedar_stages}. Node A starts from the raw SigMF frame, and each subsequent node resumes from the output of the preceding group. Cut points are allowed after stages 3--13, and the identity, head, and post-processing stages run together on the last node, which returns the detections to Node A. On our four-node testbed, a partition therefore chooses three of these eleven cut points, which gives 165 candidate partitions for each model and bandwidth environment. For a given policy, every compressible boundary uses the same compressor, and the receiver reconstructs its input before executing its stages, so partition selection accounts for both the computation moved to another node and the representation sent across the new boundary.

\begin{table}[!ht]
\centering
\caption{Profiled stages of \cedaryolo{} and the data sent when a cut point follows each stage.}
\label{tab:app_cedar_stages}
\begin{tabular}{cllcc}
\toprule
Stage & Operation & Output & Cut allowed & Compressed \\
\midrule
1 & Read I/Q samples & Complex I/Q vector & -- & -- \\
2 & Spectrogram & Magnitude spectrogram & -- & -- \\
3 & Frequency shift and logarithm & Log-magnitude spectrogram & \(\checkmark\) & -- \\
4 & Rendering limits & Spectrogram with rendering limits & \(\checkmark\) & -- \\
5 & Render & RGBA image & \(\checkmark\) & -- \\
6 & Convert to RGB & RGB image & \(\checkmark\) & -- \\
7 & Build input tensor & Input tensor & \(\checkmark\) & \(\checkmark\) \\
8--12 & Backbone blocks 1--5 & Feature tensor & \(\checkmark\) & \(\checkmark\) \\
13 & Asymmetric SPPF & Feature tensor & \(\checkmark\) & \(\checkmark\) \\
14 & Identity (kept by the model interface) & Feature tensor & -- & -- \\
15 & Detection head & Raw predictions & -- & -- \\
16 & Post-processing & Detections & Return to Node A & -- \\
\bottomrule
\end{tabular}
\end{table}

\emph{Computation profiling.} The stage compute times \(t_s\) and boundary sizes \(b_\ell\) in~\eqref{eq:cedar_partition_optimization} come from a local run of each model on the test set. For each stage, we sum the times of its operations within each frame and take the median over frames, and the boundary size is the median output size. Because the four Jetson nodes are homogeneous, one compute profile serves every node.

\emph{Communication profiling.} The link rates \(B_\ell\) and compression ratios \(r_\ell\) come from a preliminary distributed sweep with 100 frames per setting. Its cut points, after stages (3,4,5), (4,5,9), and (7,8,9) for the nano, small, and extra-large models, expose the preprocessing outputs, the input tensor, and the early backbone activations. For each directed link and bandwidth environment, the link rate is the median of payload size over send time, pooled over models and compressors, and the return link is profiled separately. Each compressed payload is matched to its uncompressed reference, and ratios of at least 0.98 are excluded to separate effective compression from transfers whose payload is unchanged. We take the median ratio per compressor and payload type for the input tensor and the first two backbone outputs, and other boundaries use the pooled median ratios: 0.25 for INT8, 0.13 for INT4, 0.28, 0.53, and 0.78 for top-k with 25\%, 50\%, and 75\% retention, and 0.29 for the hybrid compressor, as also shown in the legend of Fig.~\ref{fig:cedar_yolo_mdi_comparison}. Multiplying a boundary's uncompressed size by its ratio gives its estimated compressed size.

\emph{Policy selection.} The feasible set \(\mathcal{P}\) in~\eqref{eq:cedar_partition_optimization} contains the policies that assign every stage exactly once and use runtime-supported cut points with available cost profiles. Candidates are ranked by their estimated \(\tau\), with the total compute and transfer time as the secondary criterion. Ranking is performed separately for each model, bandwidth environment, and compressor, so the selected stage assignments can change with the operating condition.

\emph{Checkpoints.} Owing to the experiment schedule, the \mdi{} experiments use \cedaryolo{} checkpoints from an earlier training run than the newer models of the parameter sweep and Fig.~\ref{fig:cedar_yolo_local_comparison}, with slightly different training settings. The newer models reach slightly higher detection quality overall. The \mdi{} experiments evaluate the inference efficiency, which does not depend on the trained weights, and the detection quality they report is that of the checkpoints they use.

\emph{Procedure.} Before each case, Node A distributes the model, partition, compressor, and link-rate configuration to the workers. The workers construct their assigned execution groups, load checkpoints where model computation is required, and acknowledge readiness, after which Node A starts the case timer. The partition and compressor stay fixed throughout the case, and the runtime drains the case before switching configurations. Overlap F1 is computed from confusion counts pooled over the full test set.

\emph{Results.} In Fig.~\ref{fig:cedar_yolo_mdi_comparison}, every run whose overlap F1 is within 0.01 of local execution exceeds local throughput while keeping the detection quality of local execution. The nano and small models reach about 1.1 frames/s, 1.19 and 1.47 times their local throughput, in every setting, independent of the bandwidth and the compressor, because the preprocessing on Node A, about 0.89~s per frame, takes longer than the computation on every other node and the transfer on every link. The extra-large model benefits most from \mdi{}, and its best gain grows with the bandwidth, from 2.23 times the local throughput in the low-bandwidth environment to 3.17 and 3.27 times in the medium- and high-bandwidth environments. The top-ranked candidate R1 is close to the best of the three candidates in every case.

\FloatBarrier
\end{document}